\documentclass[conference]{IEEEtran}
\IEEEoverridecommandlockouts
\usepackage{cite}
\usepackage{amsmath,amssymb,amsfonts}
\usepackage{algorithmic}
\usepackage{graphicx}
\usepackage{textcomp}
\usepackage{xcolor}
\def\BibTeX{{\rm B\kern-.05em{\sc i\kern-.025em b}\kern-.08em
    T\kern-.1667em\lower.7ex\hbox{E}\kern-.125emX}}
\begin{document}

\title{A Service for Supporting Digital and Immersive Cultural Experiences}

\author
{\IEEEauthorblockN{Karthik Vaidhyanathan}
\IEEEauthorblockA{\textit{University of L'Aquila}\\
karthik.vaidhyanathan@univaq.it}
\and
\IEEEauthorblockN{Antonio Bruno}
\IEEEauthorblockA{\textit{University of L'Aquila}\\
antobru28@gmail.com}
\and
\IEEEauthorblockN{Eleonora Mendola}
\IEEEauthorblockA{\textit{University of L'Aquila}\\
mendola.eleonora@gmail.com}
\and
\IEEEauthorblockN{Filippo Mignosi}
\IEEEauthorblockA{\textit{University of L'Aquila}\\
filippo.mignosi@univaq.it}\\
\and
\IEEEauthorblockN{Mahyar T. Moghaddam}
\IEEEauthorblockA{\textit{MMMI Institute - University of Southern Denmark}\\
mtmo@mmmi.sdu.dk}
\and
\IEEEauthorblockN{Henry Muccini}
\IEEEauthorblockA{\textit{University of L'Aquila}\\
henry.muccini@univaq.it}
\and
\IEEEauthorblockN{Monica Nesi}
\textit{University of L'Aquila}\\
monica.nesi@univaq.it}

\maketitle

\begin{abstract}
Cultural heritage sites in Italy typically attract a large number of tourists every year. However, the lack of support for i) locating contents of interest;  ii) discovering information on specific contents; and iii) ease of navigation within the heritage site; hinders the overall experience of the visitor. To this end, in this work, we present a Digital Object Space Management service developed as a part of the VASARI project. The service generates a digital twin (with 3D visualization) of a given cultural heritage site and further provides support for navigation and localization, thereby providing an immersive cultural experience to the visitor.
%It caters to two key stakeholders: i) heritage site managers who can upload a 2D image(s) of the museum map, mark POI's, artworks, etc. ii) visitors who will be able to use the service for indoor navigation and localization, generating customised visit routes and for locating artworks in a museum thereby providing the visitor with an immersive visiting experience.

\end{abstract}


\section{Introduction}

Italy is known for the presence of rich cultural heritage sites, attracting millions of visitors every year. As per the latest data from Statista, the cultural heritage sites in Italy have attracted around 67 million visitors in the last two years~\cite{statista}. Although this is the case, the cultural tourism market is evolving towards a dimension of complete satisfaction of the needs of the tourist, enhancing, on the one hand, the centrality of the cultural aspect using 360° experiences and on the other hand by creating customized visit routes. The visitors are also showing a growing interest to play an active and participatory role in the tourism experience, integrating the cultural content of the visit with self-generated personal content and sharing them with the community. However, cultural heritage in Italy, especially museums, is often under-exploited and does not provide adequate supports for visitors in i) locating contents of interest;  ii) discovering information on specific contents; iii) navigating within the heritage sites and iv) creating customized visit routes based on interests. Thereby limiting the affecting the overall experience of the visitor.
To meet these needs, the VASARI project\footnote{VASARI  ARS01\_00456 is co-funded by MIUR, with the support of EU} proposes a novel paradigm for sustaining on-site, immersive, inclusive, and contextualized user experiences by exploiting several recent technologies, such as IoT and mobile computing, semantics, big data processing, virtual and augmented reality within an innovative infrastructure based on microservices deployed in edge, fog and cloud nodes. It follows a User-Centered Paradigm that aims to create an integrated virtual-physical space where sites, cultural assets, and users coexist in the two spaces simultaneously.

In this work, we present the Digital Object Space Management Service (DOSM) developed as a part of the VASARI project. It provides support for the generation of a digital twin~\cite{digtwin} that captures the physical space and the cultural assets. Further, it leverages IoT devices to enable navigation and localization support for visitors, thereby providing them with an immersive visiting experience. In particular, the service caters to two main stakeholders: i) The heritage site administrators/staff who can add/modify/remove a digital representation of a given physical space, add/modify or remove: the POI's, cultural assets, and the position of objects, etc. %add/modify, remove descriptive contents for/of the cultural assets, etc. 
ii) visitors who can make use of this service on their mobile apps to get real-time 3D visualization of the physical space, track their real-time locations, locate cultural assets/POI's, fetch detailed information of a cultural asset,  navigate to cultural assets/POI's, generate customized navigations routes based on interests, etc.

\section{Scenarios and Models}

We envision several engaging scenarios where 
the VASARI DOSM service could be used to provide an enhanced and immersive visiting experience. We describe here three key scenarios. The first two scenarios concern the visitor before the visit and during the visit, while the third concerns the administrator/staff of the cultural heritage site: i) A prospective visitor intends to visit a cultural heritage site, and before the visit, she may want to use a mobile/web app that provides the 3D map of the heritage site, to go through the list of assets (exhibits), read about them and mark her preferred assets; ii) The visitor arriving at the museum can make use of the app on her mobile/tablet to get a 3D immersive view of the heritage site, locate her real-time position in the heritage site, generate a navigation path that connects her preferred assets that she had marked provided prior to the visit, get notification whenever she reaches near any asset, clicking on which gets detailed information on the asset.

To support the above scenarios, the VASARI DOSM service provides mechanisms that allow physical space to coexist with digital space through the definition of views (realized at different levels of abstraction). These views provide a mapping between the physical and digital space. The entities of the physical space are the environments in the heritage site where the works are placed (e.g., museum rooms associated with floors of buildings, areas in open spaces, etc.). In effect, the service supports the generation of a {\em digital twin} of the physical space, where the physical space corresponds to the digital representation with a one-to-one relationship. In particular, the physical space of interest is represented through 3D models, and the same types of modeling are used for representing the works (cultural assets, artworks, etc.). The result will be a cyber-physical space in which the physical space is represented, enriched, and complemented by digital components. Further, the use of sensor devices (BLE beacons, RFID readers, mobile devices of the visitors, etc.) is leveraged for enabling support for {\em navigation and localization} services for visitors to provide an immersive visiting experience.

%The mapping of physical space into digital space is enabled using different sensory elements in physical space. These can range from people involved in configuring the environment to nodes electronic sensors used to observe the areas of interest or abstractions (context information) of areas of interest, such as the physico-chemical parameters that characterize the environments modeled and represented digitally (e.g. RGB / IR cameras, RFID tags and readers, BLE beacons, environmental sensors, presence, etc.). 

%built on the top of the digital object management service developed as a part of the VASARI project. The service  

\section{Digital Object Space Management Service}
The VASARI DOSM service provides two key functionalities i) Digital Twin Generation ii) Navigation and Localization. In this section, we provide high-level details on how these functionalities are accomplished by the DOSM service. 
%which allows the administrators/staff of the heritage sites to create a 3D digital model of the physical space; ii) Navigation and Localization, which allows the visitors to locate POI's, navigate within the physical space, get information on POI's as well as create customized navigation routes. 

\subsection{Digital Twin Generation}
Generation of digital twin comprises of three processes, namely: i)~\textit{Panoramic Scanning}, ii)~\textit{2D to 3D Conversion}, and iii)~\textit{POI Tagging}. \noindent \textit{Panoramic scanning} is performed with the help of a camera by choosing a starting point in the physical space. Following this, the camera is rotated to perform a 360 scan of all the observable points with respect to the starting point. The choice of the observable points needs to be done in such a manner that the time required to scan the space is optimal such that no points are left uncovered. This can be done with different types of devices such as high-resolution mobile cameras, standard digital cameras, etc. This process results in a panoramic 2D image composed of a set of 2D images that capture the 360 views of a given physical space. The 2D panoramic image generated is then converted to a 3D representation with the help of a {\em 2D to 3D Conversion} process.  The process achieves this by making use of Matterport technology\footnote{https://matterport.com/} along with an API layer provided by e-Building\footnote{https://www.e-building.it/}. The process converts the $x,y$ coordinates of the captured images into $x,y,z$ coordinates based on the panoramic scans performed. The starting point of the panoramic scan is considered as origin $(0,0,0)$ and every scan made to left or right is considered as the corresponding increase/decrease in $x,y$ respectively and the movement made towards or away from the physical space during the scan becomes the $z$. This process generates a 3D digital map of the physical space. Following this, the {\em POI Tagging} process allows heritage administrators/staff to mark the POIs or cultural assets on the generated map. This is performed using the concept of mattertags provided by Matterport. It allows the assignment of dedicated tags with unique ID along with a description for each of the cultural assets/POIs.
Further, each such tag has an associated position (x,y,z) in the 3D map of Matterport. This information is further stored in a database. The above three processes generate a complete digital representation of a physical space resulting in a digital twin.

\subsection{Navigation and Localization}
The DOSM service achieves this with the help of three main processes, namely, i)~\textit{POI Device Mapping}, ii)~\textit{Visitor Tracking} and iii)~\textit{Path Generation}. The \textit{POI Device mapping} involves mapping each of the cultural assets to an IoT device ( BLE Beacon in this work). The ID of the cultural asset, along with that of the beacon, is stored in a database. This information is used by the \textit{Visitor Tracking} process. Whenever a visitor is in the vicinity of a particular cultural asset, the service running on the visitor's device receives the signals from all the beacons. These signals are used in calculating the distance between the visitor's device and different beacons. Once these distances have been obtained, the service uses a triangulation technique to obtain the visitor's position (relative to a cultural asset). The position information is then appended on the Matterport 3D map to denote the visitor location. Further, based on this position, the cultural asset closer to the visitor is estimated, and this information is used by the service to display a detailed description of the cultural asset. The \textit{Path Generation} process involves generating the navigation path for the visitor based on her preferences by also considering her current location. This is performed by generating a navigation graph that considers the different rooms, walls, cultural assets, and other POI's. Such information can be obtained by associating a mattertag to each wall, room, etc. Further, the graph also considers the visitor's preferred cultural assets. Based on this, the process uses a 2D Dijkstra algorithm to compute the shortest path that includes the visitor's preferred assets.

%\begin{itemize}
%    \item Space Modeling 
%    \item POI Tagging
%    \item Device Mapping
%    \item Localization
%\end{itemize}

\section{Conclusions}
In this work, we presented the digital object space management service developed as a part of the VASARI project. The service provides support for generating a digital twin of a given cultural heritage site and further leverages IoT devices to provide navigation and localization support for visitors, thereby providing an immersive cultural experience.

\end{document}